\documentclass[12pt,a4paper]{article}
\usepackage{graphicx}
\usepackage{amssymb,amsmath,amstext,amsthm,amsfonts}

\def\be{\begin{equation}}
\def\ee{\end{equation}}
\def\bea{\begin{eqnarray}}
\def\eea{\end{eqnarray}}
\def\bear{\begin{array}}
\def\ear{\end{array}}
\def\bfig{\begin{figure}}
\def\efig{\end{figure}}
\def\bcen{\begin{center}}
\def\ecen{\end{center}}
\def\raw{\rightarrow}

\def\Bra#1{\bigl\langle #1\bigr|}
\def\Ket#1{\bigl| #1\bigr\rangle}

\def\vq{\mathbf{q}}

\def\chic{\scriptscriptstyle}

\def\bi{\begin{itemize}}
\def\ei{\end{itemize}}

\begin{document}

\title{\bf On the nature of the Roper resonance}
\author{L. Alvarez-Ruso\\Centro de F\'{\i}sica Computacional, Departamento de F\'{\i}sica,\\ Universidade de Coimbra, Portugal}
\date{November 2, 2010}
\maketitle

\begin{abstract}
The lightest $N^*$ state, $N(1440)$~$P_{11}$, also known as Roper resonance, has puzzled physicists for decades. A large variety of theoretical models aimed to understand its properties have been proposed. Some of them are briefly reviewed here, together with the hadronic processes where the Roper resonance is revealed or plays an important role. 
\end{abstract}

\section{Roper resonance properties} 

In the 1950ies, Fermi and coworkers started to measure pion-nucleon cross sections and to analyze the data in terms of partial waves, leading the way to the discovery of a large number of baryon resonances. In 1963, in a partial-wave analysis performed at the Lawrence Livermore National Laboratory,  L. D. Roper found a $P_{11}$ resonance at    
$\sqrt{s} \approx 1.43$~GeV ($\approx 600$~MeV pion laboratory kinetic energy)~\cite{Roper:1964zz}. The result was surprizing as there were no hints for such a state and the $P_{11}$ scattering length is rather large and negative. In words of Roper: {\it I spent a much time trying to eliminate the $P_{11}$ resonance}~\cite{Ropersite}. 

The Particle Data Group estimates for the main $N^*(1440)$ properties are listed in Table~\ref{tab1}. 
\begin{table}[h!]
\begin{center}
\begin{tabular}{l r}
\hline 
\multicolumn{2}{c}{$N(1440)$~$P_{11} \qquad \qquad I(J^P)=1/2(1/2^+)$} \\
\hline
\multicolumn{2}{l}{Breit-Wigner mass = 1420 to 1470 ($\approx 1440$)~MeV} \\[0.05cm]
\multicolumn{2}{l}{Breit-Wigner full width = 200 to 450 ($\approx  300$) MeV} \\ [0.05cm]
\multicolumn{2}{l}{Re(pole position) = 1350 to 1380 ($\approx  1365$) MeV } \\[0.05cm]
\multicolumn{2}{l}{−2Im(pole position) = 160 to 220 ($\approx  190$) MeV } \\ [0.2cm]
Decay modes & Fraction ($\Gamma_i/\Gamma_{\mathrm{tot}}$) \\
\hline 
$N \pi$ & 0.55 to 0.75 \\[0.05cm]
$N \pi \pi$ & $30 - 40$~\% \\[0.05cm]
$\qquad \Delta \pi$  & $20 - 30$~\% \\[0.05cm]
$\qquad N \rho$ & $<$ 8~\% \\[0.05cm]
$\qquad N (\pi \pi)^{I=0}_{S-wave}$ & $5 - 10$~\% \\[0.05cm]
$p \gamma$ & $0.035 - 0.048$~\% \\[0.05cm]
$n \gamma$ & $0.009 - 0.032$~\% \\
\hline
\end{tabular}
\caption{Summary of the PDG estimates for the Roper resonance properties~\cite{Amsler:2008zzb}.} 
\label{tab1}
\end{center}
\end{table}
Considerable uncertainties are apparent, specially in the full Breit-Wigner width and the branching ratios to the strong-decay channels. Indeed, different values are obtained with different models, most of them built in terms of Breit-Wigner resonances plus background, meson-exchange or $K$-matrix formalisms. For example, the recent $K$-matrix multichannel analysis of Ref.~\cite{arXiv:0707.3591}, which combines single and double-pion production data induced by pions and photons finds a $\Gamma_{\pi N}/\Gamma_{\mathrm{tot}} \approx 61$~\%, in agreement with the PDG, but a smaller $\Gamma_{\pi \Delta}/\Gamma_{\mathrm{tot}} \approx 18$~\% and a considerably larger $\Gamma_{\sigma N}/\Gamma_{\mathrm{tot}} \approx 21$~\% (to be compared to the $N^* \raw N (\pi \pi)^{I=0}_{S-wave}$ 5-10~\% PDG estimate). 

Pole positions and residues allow for a parameterization of resonances in a well-defined way, free of assumptions for the background and energy dependence of the resonance part~\cite{arXiv:0903.4337}. Actually, many different studies find for the Roper resonance two almost degenerate poles close to the $\pi \Delta$ threshold on two different Riemann sheets of the $\pi \Delta$ channel~\cite{Arndt:1985vj,arXiv:0903.4337,arXiv:0909.1356,Kamano:2010ud}. The pole positions are stable against larger variations of parameters in meson-exchange mechanisms, with averaged values of $(\mathrm{Re} M^*,- \mathrm{Im} M^*) = (1363^{+9}_{-6}, 79^{+3}_{-5})$~MeV  and $(1373^{+12}_{-10}, 114^{+14}_{-9})$~MeV~\cite{Kamano:2010ud}.    
The second pole is a replica or shadow of the first one without strong physical implications rather than a new structure~\cite{arXiv:0903.4337}. In spite of this agreement, the dynamical origin of the Roper poles is not clear: while in the JLMS model of Ref.~\cite{arXiv:0909.1356}, they evolve from a single bare state that also gives rise to the $N^*(1710)$, no genuine pole term is required in the J\"ulich model~\cite{arXiv:0903.4337}.  

\section{(Some of) the many faces of the Roper resonance}

In a simple quark model with a harmonic oscillator potential it is easy to understand why it is unexpected to have a radial excitation of the nucleon as the first $N^*$. The energy spectrum is given by $E_n=\hbar \omega (n + 3/2)$ with $n=n_r + l$. If the lowest state with $n=0$, $l=0$ is associated with the nucleon ($J^P = 1/2^+$), then the first excited state with $n=1$, $l=1$ is $N^*(J^P = 1/2^-)$ and only the next one with $n=2$, $l=0$ is an $N^*(J^P = 1/2^+)$ like the Roper. However, the first negative parity state $N(1535)$~$S_{11}$ turns out to be heavier than the $N(1440)$~$P_{11}$. This parity reversal pattern cannot be described by successful quark models based on $SU(6)$ symmetry with residual color-spin interactions between quarks (see for instance Fig.~9 of Ref.~\cite{nucl-th/0008028}).  

Some authors argue that reverse parity is an indication that at low energies the interactions among constituent quarks could be dominated by flavor-dependent Goldstone boson exchange (GBE) (see Ref.~\cite{Glozman:1995fu} for a review). With this assumption it is possible to obtain a good description of the low-lying baryon spectrum and, in particular, the correct level ordering between the $N^*(1440)$ and the $N^*(1535)$, as can be seen in Fig.~4 of Ref.~\cite{Glozman:1999vd}. The model has been extended to include the exchange of vector and scalar mesons to account for correlated multiple GBE, although the special nature of pseudoscalar Goldstone bosons does not extend to other mesons. Besides,   
the special status of mesons in this model makes it difficult to achieve a unified description of both mesons and baryons~\cite{nucl-th/0008028}. 

Further understanding of the nature of the Roper resonance and the level ordering may be provided by lattice QCD. In a recent study, the first positive and negative parity excited states of the nucleon have been obtained with variational analysis in quenched QCD~\cite{Mahbub:2009aa,Mahbub:2010me}. The $1/2^-$ state is below the $1/2^+$ one for heavy quark masses, but the physical ordering is recovered for pion masses below 380~MeV (see Fig.10 of Ref.~\cite{Mahbub:2010me}). Caution should be exercised in the interpretation of this result obtained in quenched QCD and for which the identification of the $1/2^-$ at low quark masses, where finite lattice volume effects become significant, still remains. If confirmed, this level crossing could support the hypothesis that there is a transition from heavy quarks, where $SU(6)$ symmetry with color-spin interactions works well, to light quarks where flavor-spin interactions due to GBE prevail~\cite{Mathur:2003zf}.

To circumvent the parity reversal problem, alternative descriptions in which the Roper resonance is not (only) a $qqq$ state have also been proposed. For instance, it could have a large gluonic component $q^3G$, although the masses of such hybrid states calculated with the flux-tube model are quite large ($M_{hyb} > 1870 \pm 100$~MeV)~\cite{Capstick:1999qq}.
In one of its oldest representations, the Roper appears as a collective vibration of the bag surface, a breathing mode. Indeed, with the Skyrme model, where baryons are topological solitons of the meson nonlinear fields, a resonance was found in the breathing mode spectrum with a mass of $M^*=1420$~MeV~\cite{Kaulfuss:1985na}. In line with the collective picture, Juli\'a-D\'iaz and Riska explored the presence of $(q \bar{q})^n$ components in the Roper resonance~\cite{JuliaDiaz:2006av}. They found that the confining interaction mixes the $qqq$ and $qqq\bar{q}q$ components. The $qqq\bar{q}q$ admixture in the Roper ranges from 3 to 25~\% depending on the constituent quark mass while the $qqq(\bar{q}q)^2$ components are negligible. The $qqq$ component could even be totally absent in the $N^*(1440)$ as suggested by the fact that the resonance shape is dynamically generated in the J\"ulich model from meson-baryon interactions in coupled channels, mostly from the $\sigma N$ S-wave interaction~\cite{Krehl:1999km,arXiv:0903.4337}. Finally, if the baryons are regarded as many-body systems of quarks and gluons, it is natural to expect that they could be deformed. Such a possibility was investigated in Ref.~\cite{Hosaka:1998xp}, with a deformed oscillator potential. It was shown that low lying masses fit well to rotational spectra with the Roper as an $n=2$ rotational state. 

\section{Hadronic reactions} 

Although the vast majority of the information about the $N^*(1440)$ has been extracted from the $\pi N \raw \pi N$ reaction, there are many other processes where the resonance properties can be studied and/or where the reaction mechanism cannot be understood without taking it into account. Some of these processes are reviewed in this Section. 

\subsection{Electroproduction of the $N^*(1440)$}

Valuable information about nucleon resonances is encoded in the electromagnetic $N \raw N^*$ transitions, often presented in terms of helicity amplitudes connecting states with well defined helicities. In the case of the $N-N^*(1440)$ transition, two such amplitudes should be introduced,  $A_{1/2}$ and $S_{1/2}$, defined as
\begin{eqnarray}
A_{1/2}(q^2) & = & \sqrt{\frac{2 \pi \alpha}{k_R}} {\Bra{{N^*
\downarrow}}} \epsilon^{\chic (+)}_\mu 
{ J^{\mu}} {\Ket{{ N \uparrow}}} \,, \\
S_{1/2}(q^2) & = & \sqrt{\frac{2 \pi \alpha}{k_R}}
\frac{|{\mathbf{q}|}}{\sqrt{-q^2}}
{\Bra{{ N^*\uparrow}}} \epsilon^{\chic (0)}_\mu 
{J^{\mu}} {\Ket{{N \uparrow}}} \,.
\end{eqnarray}
Here, $\alpha$ is fine-structure constant, $k_R=(M^2_{\chic{N^*}} - M^2_{\chic{N}})/ (2 M_{\chic{N^*}})$, $q=(\omega, \vq)$ is the four-momentum transfered to the nucleon and $\epsilon^{\chic (+,0)}$ stand for the transverse and longitudinal polarizations of the virtual photon. The $N-N^*(1440)$ transition electromagnetic current can be parametrized with two form factors 
\begin{equation}
J^\mu  = \bar u_{N^*}(p') \left[ {F_1}(q^2)\left( q \!\!\!\!\: /\,
q^{\mu} -
q^2 \gamma^{\mu}\right) + i {F_2}(q^2) \sigma^{\mu \nu} q_{\nu}
\right] u(p) \,.
\label{current}
\end{equation}
This current is very similar to the nucleon one, except for the $ q \!\!\!\: /\, q^{\mu}$ part. In the nucleon case, the form factor associated with this operator has to
vanish to ensure current conservation, but not for the $N-N^*$ transition because the Roper mass differs from the nucleon one. Introducing electric and magnetic form factors, in analogy to the Sachs form factors of the nucleon and
substituting Eq.~(\ref{current}) in the expressions for the helicity amplitudes, one obtains that up to well known factors $A_{1/2} \sim  {G_{\chic{M}}}$ and    
$S_{1/2} \sim  {G_{\chic{E}}}$~\cite{Cardarelli:1996vn,AlvarezRuso:2003gj}.  

\begin{figure}[ht]
\begin{center}
\includegraphics[width=0.65\textwidth]{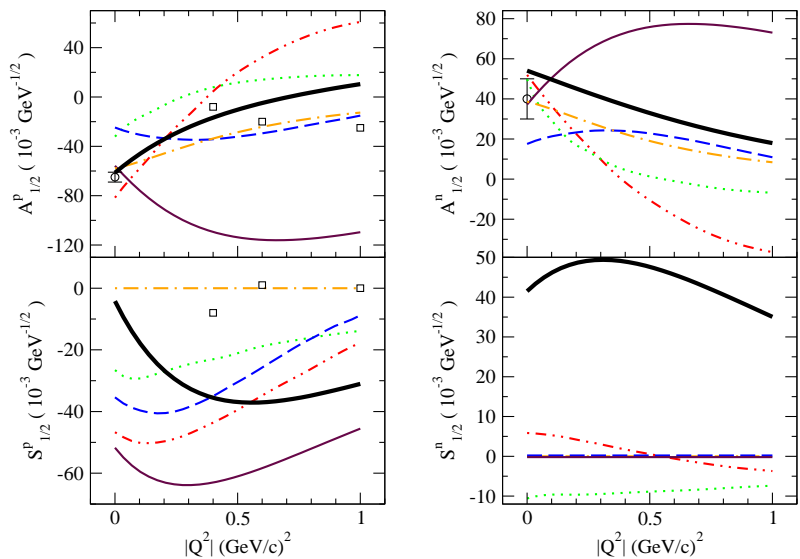} 
\caption{Transverse ($A_{1/2}$) and longitudinal ($S_{1/2}$) helicity amplitudes for the $N-N^*(1440)$ transition calculated with various models: NRQM (solid line)~\cite{Li:1991yba}, hybrid model (dash-dotted line)~\cite{Li:1991yba}, LF (dotted line)~\cite{Cardarelli:1996vn}, ChD (dashed line)~\cite{Alberto:2001fy} and EVMD (dash-double-dotted line)~\cite{nucl-th/9804071}. The result of the global MAID07 analysis~\cite{arXiv:0710.0306} is given by the thick solid line.}
\label{hel}
\end{center}
\end{figure}

The $N-N^*(1440)$ helicity amplitudes have been studied using various models
with a wide diversity of results. Some of these are shown in Fig.~\ref{hel}, namely,
the prediction from the non-relativistic quark model (NRQM)~\cite{Li:1991yba}, the hybrid
model~\cite{Li:1991yba}, the light-front relativistic quark model (LF) calculation of Ref.~\cite{Cardarelli:1996vn}, the chiral chromodielectric (ChD) model~\cite{Alberto:2001fy} and
the extended vector-meson dominance (EVMD) model of Ref.~\cite{nucl-th/9804071}.

The extensive $N^*$ program at JLab has provided a large amount of precision data on pion electroproduction which, together with the data from previous experiments at MIT/Bates and MAMI/Mainz, has made possible the extraction of the transition helicity amplitudes at $0 < Q^2 < 6$~$(\mathrm{GeV}/c)^2$  the for several resonances and, in particular, for the Roper~\cite{arXiv:0710.0306,arXiv:0909.2349}.  The result from the global MAID07 analysis is also shown in Fig.~\ref{hel}. The comparison with the models reveals that none of them is really satisfactory. This is an indication of the difficulties that quark models encounter in the description of the low $Q^2 <1$~$(\mathrm{GeV}/c)^2$ region. At $Q^2 > 2$~ $(\mathrm{GeV}/c)^2$, where $A^p_{1/2}$ and  $S^p_{1/2}$ are positive and decreasing, good agreement is obtained with relativistic quark model calculations assuming that the Roper is the first radial excitation of the nucleon~\cite{arXiv:0909.2349,Ramalho:2010js}. The discrepancies at low $Q^2$ are interpreted as due to the missing meson cloud effects. The importance of the pion cloud, particularly at low $Q^2$, has also been demonstrated in a recent study of electroproduction amplitudes with the simple Cloudy Bag Model~\cite{arXiv:0906.2066}. The pion cloud is found to be responsible for the large and negative value of $A^p_{1/2}$ at the photon point, while the quark dynamics becomes progressively relevant as $Q^2$ increases, causing $A^p_{1/2}$ to change sign.

It is important to bare in mind that extraction of helicities amplitudes in both the MAID~\cite{arXiv:0710.0306} and CLAS~\cite{arXiv:0909.2349} analyses imply certain model dependent assumptions about the resonant and non-resonant parts of the pion electroproduction amplitudes. For this reason, alternative methods are being pursued, like the extraction of transition form factors at the resonance poles using analytic continuation~\cite{arXiv:1006.2196}. 

\subsection{Direct observation of the Roper resonance}

The excitation of the Roper resonance in $\pi N$ and $\gamma N$ reactions can only be assessed with partial wave analyses; in the reaction cross section, the $N(1440)$~$P_{11}$ overlaps with the  $N(1520)$~$D_{13}$ and the $N(1535)$~$S_{11}$ forming the so called second resonance region. Moreover, all these $N^*$ states might be masked by the prominent $\Delta(1232)P_{33}$ excitation since $\pi N$ and $\gamma N$ interactions mix isospin 1/2 and 3/2. However, certain reactions act as filters, making the direct observation of the Roper excitation possible. 

An example is the $(\alpha,\alpha')$ reaction of proton target studied at SATURNE with a beam energy of 4.2~GeV~\cite{Morsch:1992vj}. As the projectile has $I=0$, the $\Delta(1232)$ excitation can occur on the projectile but not on the target. For this reason the Roper excitation appears as small peak on the tail of the dominant $\Delta(1232)$ excitation (see Fig.~2 of Ref.~\cite{Morsch:1992vj}). The theoretical study of Hirenzaki et al.~\cite{Hirenzaki:1995js} showed that the isoscalar excitation on the proton is the dominant $N^*(1440)$ production mechanism and extracted its strength from data. The fact that the interference with the $\Delta(1232)$ excitation on the $\alpha$ is important allowed to establish also the relative sign of the amplitudes.  
  
An even clearer case of direct $N^*(1440)$ observation has been made by the BES Collaboration with the decay $J/\psi \raw \bar{N} N \pi$~\cite{Ablikim:2004ug}. Here, because of isospin conservation, the $\pi N$ system is in pure isospin $1/2$. Several $N^*$ were observed in the $\pi N$ invariant mass distribution, the first of them corresponding to the Roper resonance. Its mass and width, estimated with a simple Breit-Wigner function were found to be $1358 \pm 6 \pm 16$~MeV and $179 \pm 26 \pm 50$~MeV respectively. As a constant width was used in the Breit-Wigner, the extracted mass is close to the pole value.

\subsection{Double-pion production reactions}

The Roper resonance is a vital ingredient in double-pion production reaction mechanisms. In spite of its small branching ratio, the S-wave character of the $N^*(1440) \raw N (\pi \pi)^{I=0}_{S-wave}$ decay (or $N^*(1440) \raw N \sigma$ as often denoted in the literature) makes it a very important nonvanishing contribution at threshold. This is the case for the $\pi N \raw \pi \pi N$ reaction, as was shown long ago in Ref.~\cite{Oset:1985wt} and supported by other models. For instance, in Fig.~10 of Ref.~\cite{Kamano:2006vm} the dotted lines denoting the results without $N^*(1440)$ are well below the full model (and the data) in the channels where the   $N^*(1440) \raw N (\pi \pi)^{I=0}_{S-wave}$ decay is allowed.    

The relevance of the Roper is even more dramatic in $N N \raw N N \pi \pi$, where according to the model of Ref.~\cite{AlvarezRuso:1997mx}, the isoscalar excitation of the resonance, followed by its decay into $N (\pi \pi)^{I=0}_{S-wave}$ appears to be dominant at laboratory kinetic energies of the incident proton $T_p < 1$~GeV. The other two important reaction mechanisms: $\Delta \Delta$ excitation and $ N^*(1440) \raw \Delta \pi$ are negligible at threshold but rise fast to become important above $T_p = 1$~GeV. In recent years, this reaction has been accurately measured at CELSIUS and COSY. At low energies, the main features predicted by the model of Ref.~\cite{AlvarezRuso:1997mx} have been confirmed (see for instance Ref.~\cite{Patzold:2003tr}). The situation is more involved at higher energies: an isospin analysis of the data~\cite{Skorodko:2009ys} indicates that the contribution from heavier $\Delta$ states might be important. Resonances with masses up to 1.72~GeV have been incorporated in the relativistic model of Cao et al.~\cite{Cao:2010km}, finding large contributions from the $\Delta(1600)$ and $\Delta(1620)$ states. The agreement to data is improved by reducing the $N^*(1440) \raw \Delta \pi$ branching ratio, in line with the findings of Ref.~\cite{arXiv:0707.3591}. 

\begin{figure}[h!]
\begin{center}
\includegraphics[width=0.64\textwidth]{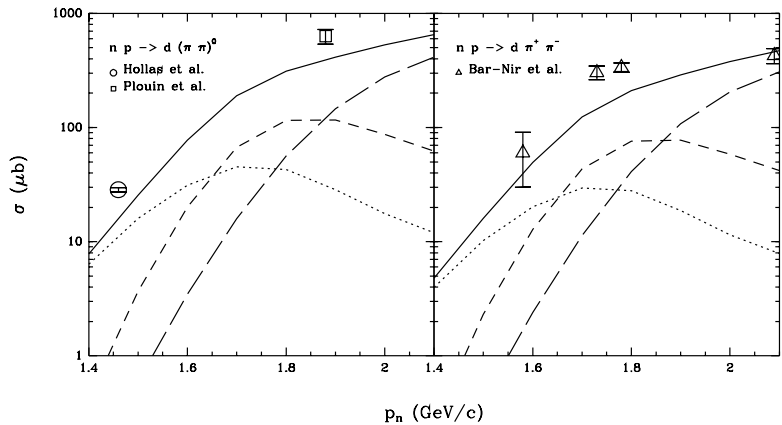}
\caption{Total cross section for $n p \raw d \pi \pi$ as a function of the neutron laboratory momentum (solid line). The dotted line corresponds to the $N^*(1440) \raw N (\pi \pi)^{I=0}_{S-wave}$ mechanism, the short-dashed line stands for the $N^*(1440) \raw \Delta \pi$ and the long-dashed one for the double-$\Delta$ excitation (see Ref.~\cite{AlvarezRuso:1998xg} for details). The data are from Refs.~\cite{Hollas:1982xs} (circle), \cite{Plouin:1978hr} (square) and \cite{BarNir:1973wb} (triangles).}
\label{deut}
\end{center}
\end{figure}

The $N N \raw N N \pi \pi$ model of Ref.~\cite{Cao:2010km} does not include interferences but, in particular, the interference between the $N (\pi \pi)^{I=0}_{S-wave}$ and $\Delta \pi$ decay modes of the Roper has been found to explain some details of the invariant mass and angular distributions for $\pi N \raw \pi \pi N$ (Fig.~12 of Ref.~\cite{Kamano:2006vm}), $N N \raw N N \pi \pi$ (Fig.~4 of Ref.~\cite{Patzold:2003tr}) and specially $n p \raw d \pi \pi$. For this later reaction, it has been shown that the shape of the double differential cross sections measured at LAMPF with  a neutron beam of $p_n = 1.463$~GeV$/c$~\cite{Hollas:1982xs} can be explained by the above mentioned interference between the two-pion decay modes of the Roper resonance~\cite{AlvarezRuso:1998xg}. As shown in Fig~\ref{deut}, by taking into account the Roper one obtains a good description of the size and energy dependence of the total $n p \raw d \pi \pi$ cross section even with a rather simple model as the one of Ref.~\cite{AlvarezRuso:1998xg}. The $n p \raw d \pi \pi$ reaction close to threshold has been recently investigated in the framework of chiral perturbation theory~\cite{Liu:2010zy}. The reported results for the total cross section are considerably smaller than those of Fig.~\ref{deut} even at lower energies.

I thank Michael D\"oring for reading the manuscript and providing useful comments.

\end{document}